\documentclass[review]{elsarticle}

\usepackage{lineno,hyperref}
\usepackage{graphicx}
\usepackage{float}
\usepackage{amssymb}
\modulolinenumbers[5]
\usepackage{multirow}
\usepackage{subcaption}
\usepackage[T1]{fontenc}
\usepackage[utf8]{inputenc}
\usepackage[numbers]{natbib}
\usepackage{pdfpages}

\journal{Information Sciences}
\begin{document}

%\definecolor{lightblue}{HTML}{72B4D7}

\begin{frontmatter}

\title{Predicting complex user behavior from CDR based social networks}
%\tnotetext[title_note_label]{title note here.}

%% Affiliations in footnotes:
\author[nest,phys]{Casey Doyle}
\author[jsi,ips]{Zala Herga}
\author[nest,phys]{Stephen Dipple}
\author[nest,cs,wust]{Boleslaw K. Szymanski\corref{ca}}
\ead{szymab@rpi.edu}
\cortext[ca]{Corresponding author}
\author[nest,phys]{Gyorgy Korniss}
\author[jsi,ips]{Dunja Mladeni\'c}

\address[nest]{Network Science and Technology Center, Rensselaer Polytechnic Institute, 110 8th Street,Troy, NY 12180, USA}
\address[phys]{Department of Physics, Applied Physics, and Astronomy, Rensselaer Polytechnic Institute, 110 8th Street,Troy, NY 12180, USA}
\address[jsi]{Artificial Intelligence Laboratory, Jo\v zef Stefan Institute, Jamova 39, Ljubljana, Slovenia}
\address[ips]{Jo\v zef Stefan International Postgraduate School, Jamova 39, Ljubljana, Slovenia}
\address[cs]{Department of Computer Science, Rensselaer Polytechnic Institute, 110 8th Street,Troy, NY 12180, USA}
\address[wust]{Faculty of Computer Science and Management, Wroc\l{}aw University of Science and Technology, Wroc\l{}aw, Poland}

\begin{abstract}
Call Detail Record (CDR) datasets provide enough information about personal interactions of cell phone service customers to enable building detailed social networks. We take one such dataset and create a realistic social network to predict which customer will default on payments for the phone services, a complex behavior combining social, economic, and legal considerations. After extracting a large feature set from this network, we find that each feature poorly correlates with the default status. Hence, we develop a sophisticated model to enable reliable predictions. Our main contribution is a methodology for building complex behavior models from very large sets of diverse features and using different methods to choose those features that perform best for the final model. This approach enables us to identify the most efficient features for our problem which, unexpectedly, are based on the number of unique users with whom the given user communicates around the Christmas and New Year's Eve holidays. In general, features based on the number of close ties maintained by a user perform better than others. Our resulting models significantly outperform the methods currently published in the literature. The paper contributes also a systematic analysis of properties of the network derived from CDR.

\end{abstract}


\end{frontmatter}

\section{Introduction}
Call Detail Record (CDR) datasets, created from cell phone logs of large groups of people, have become common in studying human behavior thanks to the large amount of detailed data they provide~\cite{Blondel2015}. These datasets typically include both basic information about the users (age, gender, location) and records of calls and text messages including the time, location, and direction of each communication. 

Here, we analyze such a dataset and focus on the properties of the underlying social network that can be obtained from the data. First, we discuss various methods for building the network, aiming to mitigate the noise inherent within the cell phone records and address other issues identified in prior work such as the definitions of links, communities, reciprocity, and data types suitable for the study. These considerations are particularly relevant to our application due to the large amounts of noise and potential biases inherent to various network representation schemes~\cite{Lambiotte2008,Ling2012,Onnela2007,Li2014,dyad_recip}. In this paper, we combine techniques based on geographic and usage features along with higher level social network measures such as centrality and community structure to predict the probability of customers defaulting on their accounts.

The specific CDR dataset used here includes the detailed call history of $500,000$ clients of a cell phone company over a three month period. 
The dataset contains also information about the users' basic demographic data (age, home district, gender, and default status at the end of the three month period) as well as usage information (frequency and duration of calls, messages, and movement records based on frequently used cell towers). 
We perform traditional network analysis on the data to create new complex features.  
We analyze weighted links based on the number of communications sent between individuals to reveal the paths between individuals, reciprocity imbalances among users, and community structure in the network.

The default status (an indicator of whether the client stopped paying the phone bill over the course of data collection) is the predicted variable for this study. It is typically accessible as a part of the user's phone records, yet it still measures a complex user behavior combining many different behavioral factors.

To encompass the wide variety of possible correlations between individual attributes, behavioral indicators as well as network metrics and the probability of default, we create an extremely broad feature set, aggregating thousands of features of varying complexity from all the facets of information contained in the CDR dataset. Then, we perform feature contribution analysis and choose the features with the strongest predictive power to balance model performance and its complexity. As discussed in the next section, this general methodology for building complex behavior models
%by creating very large sets of diverse features and then systematically choosing the best performing of them to balance the final model for complexity and performance. This
is a novel approach and our contribution to the state of the art.

\section{Related work on modeling complex human behavior}

\subsection{Mobile phone data}
Call detail records (CDR) is a standard dataset collected by telecommunication operators. For each user, it contains information about telecommunication events in which this user was involved. In recent years, it has become a popular source of information for users' behavioral analytics. 

The closest to our goals is the work on modeling credit defaults \cite{san2015mobiscore}
by building a model of the user's financial risk which yields a score that can be interpreted as the probability of default. This model outperforms the Credit Bureau scores by using thousands of weak predictors derived from CDR and demographic data.
This work uses only $60,000$ users and uses only basic level network features such as degree. 

\subsection{Credit risk management}
Our paper focuses on credit risk, which refers to the clients who may stop paying back their loans, e.g., mortgage loan, credit card spending or, in our case, cell phone bill. 
Such events are called defaults. 
Banks and companies traditionally tackle this problem by calculating credit scores or probability of default for each of the potential clients.
For individual customers data sets used to predict a customer's probability of default include demographic data, loan and credit information \cite{bae2015personal}, social media \cite{ge2017predicting,zhang2009personal} or mobile phone data \cite{bjorkegren2017behavior}.  

Traditionally, the logistic model was often used to predict defaults and today is still
useful for benchmarking thanks to its simplicity, interpretability and dependability \cite{wang2018customer,bae2015personal,kuznetsova2017modeling}. However, more sophisticated and innovative approaches are also used, like neural networks \cite{wang2018customer,zhang2009personal,kvamme2018predicting}, smart ubiquitous data mining \cite{bae2015personal}, theory of three-way decisions \cite{maldonado2018credit}, and theory of survival \cite{kuznetsova2017modeling}.  In our paper, we introduce a novel approach that starts with creating a large number of features (over 6,000 here) and then reducing them to a few well performing subsets. 

\section{Network creation and analysis}
\label{sec_network}
In this section, we define the node-level location and communication activity features that indicate how embedded the node is in the network which is likely to determine the social cost of leaving the network. 
Such social network analysis is common in working with CDR datasets~\cite{Blondel2015}, but the highly detailed information contained in the CDR comes with a large amount of noise. Quantifying what level of communication between individuals indicates a connection is challenging. This issue is made worse considering the potential bias introduced by specific patterns of communication, as generational and cultural divides are prominent in phone usage~\cite{Ling2012}. Therefore
use of case-specific methods is common. Some attempts at a more general solution to this problem include reciprocity or activity requirements for links, but these solutions suffer from losing many fine details of the system~\cite{Lambiotte2008}. More complete results can be obtained by using statistical methods to detect and remove links that are more likely to be random, but the methods come with increased computation cost~\cite{Li2014}.

In this study for detailed analysis, we use directed graphs in order to preserve the imbalances that tend to arise even among reciprocal relationships~\cite{Kovanen2011}.
We also use communication frequency between individuals to define edges. 
We primarily use a weighted network for network analysis and feature generation as it better represents relationship strength and network location properties, but the unweighted network is very useful for understanding many of other interesting properties of the network unrelated to predicting user defaults. 
%We define the edge weights using the frequency of communications between individuals to preserve the benefits of the unweighted cutoff scheme while providing a more complete view of the data.
For more information on the construction, behavior, and dynamics of unweighted graphs of this network see
Supplementary Material Sec. 1.

\subsection{Weighted network based on event frequency}

\begin{figure}[t!]
\centering
\includegraphics[width=1\textwidth]{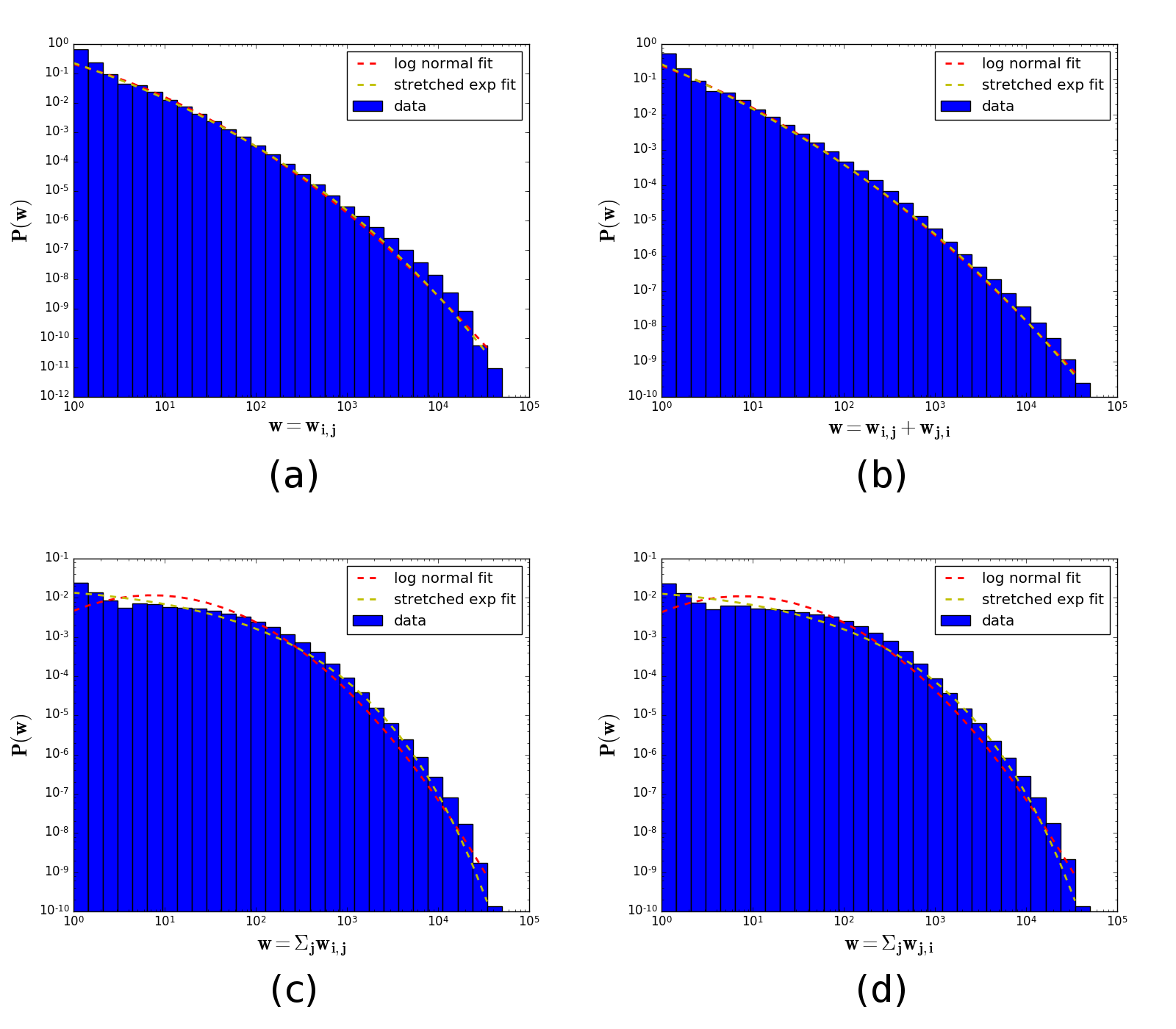}
\caption{a) Distribution of directional link weights $w_{i,j}$. (b) Distribution of non-directional link weight $w_{i,j}+w_{j,i}$. (c) Distribution of the sum of outgoing link weights of users in the system $\sum_jw_{i,j}$. (d) Distribution of the sum of incoming link weights of users in the system $\sum_jw_{j,i}$. Table \ref{tab:weight_params} shows relevant fit parameters and R-squared values. Both functions in this case reasonably fit the distributions with neither significantly outperforming the other.}
\label{fig:weight distribution}
\end{figure}

Let $w_{i,j}$ denote the number of communications sent from user $i$ to user $j$ is $w_{ij}$. We choose here the most dense representation of the network in which as long as $w_{i,j}>0$, a directed link is formed from node $i$ to node $j$ with weight $w_{i,j}$ (some interesting properties of this type of network is discussed in \cite{Yan2018,dyad_recip,Onnela2007}).
Figure \ref{fig:weight distribution} shows various probability distributions associated with $w$.
These distributions possess a curvature somewhere between an exponential distribution and a power law distribution.
It is then appropriate to fit them using a log-normal distribution and a stretched exponential which have the following form.
\begin{equation}
P(w)_{\textrm{log-normal}} \propto \frac{1}{w} e^{-\frac{(\ln w-\mu)^2}{2\sigma^2}}
\label{log_normal}
\end{equation}
\begin{equation}
P(w)_{\textrm{stretched-exp}} \propto e^{-w^\beta/\alpha}
\label{stretched_exp}
\end{equation}
\begin{table*}[t]
  \centering
  \begin{tabular}{ |c|c|c|c|c| } 
     \hline
      & $w_{i,j}$ & $w_{i,j}+w_{j,i}$ & $\sum_jw_{i,j}$ & $\sum_jw_{j,i}$ \\
     \hline
     $\sigma$ & 2.01  & 2.37 & 1.47 & 1.47 \\
     \hline
     $\beta$ & 0.127 & 0.0959 & 0.332 & 0.333 \\
     \hline
     $R^2$ log-normal & 0.9967 & 0.9991 & 0.9926 & 0.9924 \\
     \hline
     $R^2$ stretched exp & 0.9978 & 0.9994 & 0.9984 & 0.9980 \\
     \hline
  \end{tabular}
  \caption{Fitting parameters and R-squared values for Figure \ref{fig:weight distribution}. $\sigma$ corresponds to the fitting parameter in equation \ref{log_normal}. $\beta$ corresponds to the fitting parameter in equation \ref{stretched_exp}.}
  \label{tab:weight_params}
  \vspace{-0.3cm}
\end{table*}
Table \ref{tab:weight_params} shows the fitting parameters and R-squared values for Figure \ref{fig:weight distribution}.
Both fits yield high R-squared values.

We use centrality measures, reciprocity measures, and community detection to define features indicative of how embedded a user is in the network.
For all distance-based applications, we use a link's weight to define a normalized distance from the source node $i$ to the target node $j$, defined as $d_{i,j}=w_{avg} / w_{i,j}$ where $w_{avg}$ is the average weight of all connections in the network. 
To provide a baseline for comparison, we rewire the network by swapping the edge destinations to create a pseudo-random weighted graphs that maintain the in and out-degree structure of the original network.

We first look at the measure of harmonic closeness centrality (closeness centrality adapted to non-connected graphs)~\cite{rochat2009,Opsahl2010}. We define this centrality as $C_H(i)=\frac{1}{N-1}\sum_{j\neq i}\frac{1}{l_{i,j}}$, where $N$ is the total number of nodes and $l_{i,j}$ is the distance of the shortest path between nodes $i$ and $j$. For the original network, these centrality scores are generally fairly high and evenly distributed, with an average harmonic centrality of $C^{cell}_{avg}=4.61$ and a standard deviation of $C^{cell}_{std}=1.84$. Both values are higher than those for the randomly rewired graph which yields $C^{rand}_{avg}=4.11$ and $C^{rand}_{std}=1.20$. Interestingly, the diameter and average shortest path length of the giant component in the original network $D^{cell}=6.24$ and $\langle l^{cell}\rangle=.311$ are also slightly higher than the corresponding values $D^{rand}=4.35$ and $\langle l^{rand}\rangle =.30$ for the randomly rewired network. 

These differences reveal the basic shape of the original network, which is characterized by significant populations of both highly connected and highly isolated nodes. 
This property is shown clearly via the diameters of the two networks. 
The original network is significantly wider than its randomly rewired counterpart.
Despite the existence of close communities and hubs within the original network, there are multiple extremely remote nodes in it with no close ties.

We also use this weighted network to measure reciprocity~\cite{dyad_recip}, which is used to measure how one sided interactions are.  
Our first measure is how many pairwise communications are matched with communications in the opposite direction. A surprisingly low number of communications, $62.88\%$, have communications in both directions indicating that over a third of interactions are unreciprocated. 

Next, we construct a reciprocity metric that measures the average contribution of a nodes' links so that it is independent of the degree of the node. Unlike \cite{dyad_recip}, we apply it to directional links as opposed to ~\cite{dyad_recip}. 
The corresponding definition is $R_i=\frac{1}{k_i} \sum_{j\in N(i)} \frac{w_{i,j}-w_{j,i}}{w_{i,j}+w_{j,i}}$, where $k_i$ is the total (both in and out) degree of node $i$, $w_{i,j}$ is the number of communications from node $i$ to node $j$, and $N(i)$ is the set of neighbors that have a link connected to node $i$. 
According to this definition, links over which a node is sending more than receiving contribute positively to the node's metric, while links with the reverse pattern contribute negatively. 
The absolute value of the contribution itself increases monotonically with the difference in the level of communication~\cite{dyad_recip}. Finally, every link's contributions are normalized, thus bound within the range [-1,1].

\begin{figure}[t]
\centering
\includegraphics[width=1\textwidth]{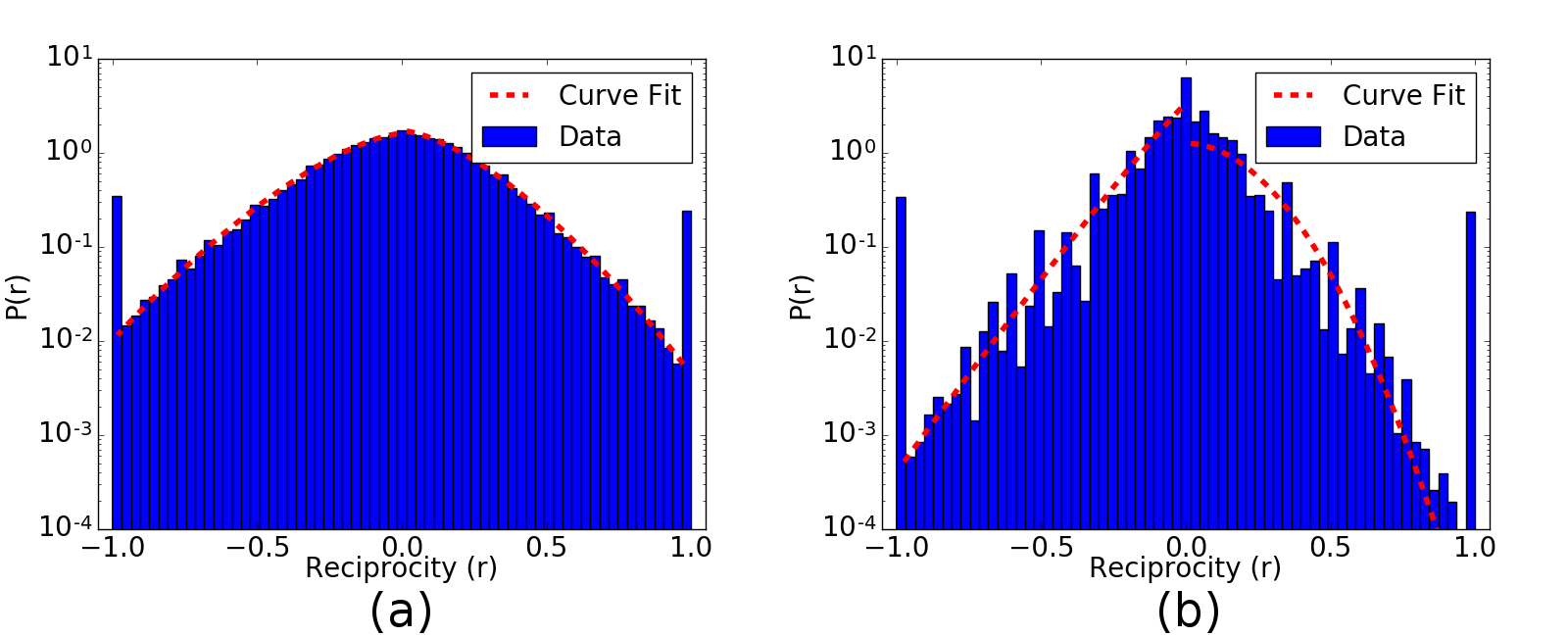}
\caption{a) The distribution of reciprocities. (b) The distribution of reciprocities using a randomized network. Peaks are observed at -1, and 1. This is consistent with a large number of nodes have low degree as most links are not reciprocated. The randomized network produces a narrower and rougher distribution indicating some level of order in how individuals interact.}
\label{fig:weighted reciprocity}
\end{figure}

We show in Fig \ref{fig:weighted reciprocity}(a) that the distribution of reciprocity is fairly smooth with large spikes at the extreme values. 
These spikes arise because there are nodes with either no incoming communications ($R_i=1$) or no outgoing communications ($R_i=-1$). 
Since all data comes from a single cell phone provider, it is likely that many of these nodes simply have contacts that use different carriers not included in the set. Removing these outliers, we fit the remaining distribution using a stretched exponential as described in Equation \ref{stretched_exp}. Table \ref{tab:weighted_recip} shows the fitting parameters for Fig \ref{fig:weighted reciprocity}. As can be seen, there is an asymmetry in the width of the positive and negative side of the distribution. This can be caused by an increased amount of communication for nodes with reciprocity equal to $1$. While each link's contribution $w_{i,j}-w_{j,i}$ is symmetric, the overall contributions are averaged and thus the symmetry can be broken. If an unreciprocated link is the only link for a node, its reciprocity will be $R_i=\pm 1$ regardless of what $w$ is. This means if there are overall more communications from nodes with $R_i=1$ than nodes with $R_i=-1$, this difference would show itself in the overall distribution. 

\begin{table*}[t]
  \centering
  \begin{tabular}{ |c|c|c|c|c| } 
     \hline 
     Fitting parameter & \multicolumn{2}{c}{CDR graph} & \multicolumn{2}{c}{Random graph} \vline \\ 
     \hline
      & Positive & Negative & Positive & Negative \\
     \hline
     $\beta$ & 1.52  & 1.49 & 1.96 & 1.07 \\
     $\alpha$ & 0.168 & 0.198 & 0.080 & 0.111 \\
     $R^2$ & 0.994 & 0.991 & 0.687 & 0.961 \\
     \hline
  \end{tabular}
  \caption{The fitting parameters used in Fig \ref{fig:weighted reciprocity}. Positive and Negative refer to the fit of the distribution where the reciprocity values are positive and negative respectively. Due to the non-integer value of the curvature, we fit the absolute value of the negative side and plot accordingly. For the CDR graph both positive and negative sides have very similar fitting parameters except for the width in the distribution which suggest an asymmetry.}
  \label{tab:weighted_recip}
  \vspace{-0.3cm}
\end{table*}

For the randomized network, Fig \ref{fig:weighted reciprocity}(b), the concavity of the distribution narrows compared to Fig \ref{fig:weighted reciprocity}(a) from the presence of more frequent values close to zero. 
Hence, the original network is more diverse than its randomized counterpart is.

For other possible representations of reciprocity within the network, see Supplementary Material Sec. 2, where we discuss two additional variants. These alternative metrics are not independent from the metric presented in this section.

\subsection{Community detection and geographical districts}
\label{communities}
\begin{figure}[t]
\centering
\includegraphics[width=1\textwidth]{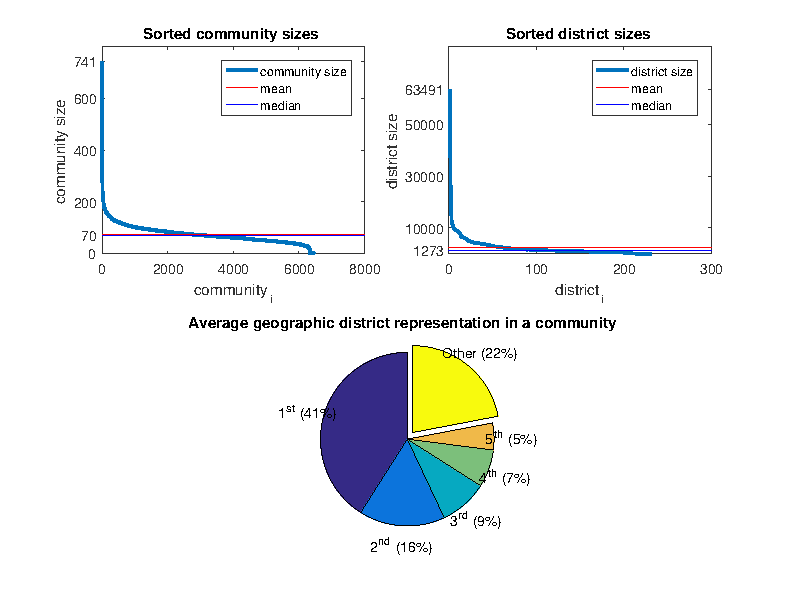}
\caption{Top: Community and districts sizes, which are sorted in descending order. The average social community size is $75$ with a median of $70$, whereas geographic districts show an average of $2,380$ and median of $1273$. Bottom: average proportion of users from a community that belong to the same geographic district - from most represented geographic district by users (\textit{$1^{st}$}) to the $5^{th}$ most represented.}
\label{fig:community_sizes}
\end{figure}

The CDR based network allows us to analyze the social communities present in the system. We use the GANXiS(SLPA) algorithm for its ability to detect even disjointed and overlapping communities and fully encapsulate the social structure of the network~\cite{Xie2011:1,Xie2011:2}.
Using this algorithm, we identify a set of over $6450$ social communities, many more than the $231$ geographic communities derived from the districts reported. 
Despite the size differences (the largest district contains $63491$ users while the largest social community has just $741$ with an average of only $74.65$ users), the groups are substantial enough to test the overlap of the lists for greater insight into how the social ties form. 
Intuitively, it seems reasonable to expect that the social communities are highly influenced by the geographic district of their members. 
Instead we see in Fig. \ref{fig:community_sizes} that on average only $41\%$ of each community comes from the same district, and in fact even the top five districts only account for $78\%$ of each community's makeup. 
The diversity of the geographic locations within social groups is especially surprising given the generally small size of the social groups compared to the geographic districts. 
The communities are also included in our feature set for predicting user default's as it is possible that individuals of a certain groups are more likely to default.

\section{Feature generation}\label{cap:features}
We combine the above described network features with the various raw usage and location features that are inherent to the dataset to create our full feature set. For ease of analysis, we divide them into six groups: 
\begin{enumerate}
\item \textbf{High Level Network features - }explained in detail above. This subset consists of $5$ features. 
\item \textbf{Consumption features - }includes information about the total number of communication events, the total and average duration of phone calls, and average time between consecutive communications. This subset consists of $2784$ features. 
\item \textbf{Correspondent features - }based on the distinct number of individuals that each user communicates with over various time periods. This subset consists of $2561$ features. 
\item \textbf{Reciprocated event features - }number of events where the observed user returns a call/message within a specified time period of receiving a communication. This subset consists of $672$ features. 
\item \textbf{Mobility features - }includes the movement patterns of individuals based on the cell tower used for each communication with relation to commonly used towers. This subset consists of $29$ features. 
\item \textbf{Location features - }includes the two most used cell phone towers for each user. This subset consists of $2$ features. 
\end{enumerate}
In some cases large or very inclusive features can be broken up by analyzing time windows (hours, days, day of the week, weeks, months, business hours, non-business hours, weekend, weekday), direction (ingoing$ / $outgoing event), and communication type (call$ / $message), leading to a large overall set of more than 6000 features. Most of these features are standard, so we do not present them in detail here, but relegate the full descriptions of each feature set to
Supplementary Material Sec. 3.

\section{Predicting default status}

\subsection{Correlation analysis}
\label{sec_correlation}
\begin{figure}[t]
\centering
\includegraphics[width=0.9\textwidth]{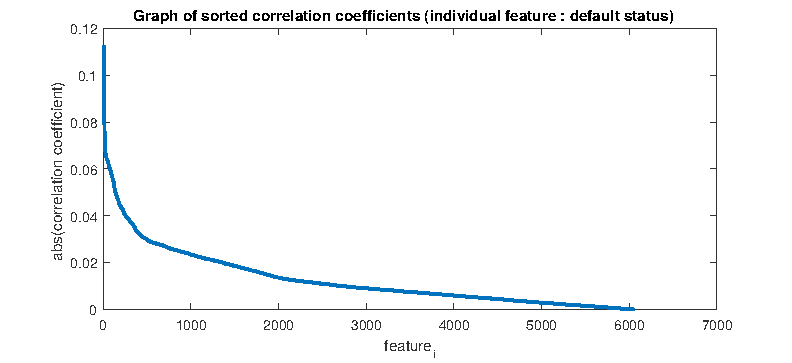}
\caption{This graph shows point biserial correlation of each individual feature to default status. Values are sorted in descending order by absolute value of their correlation coefficient. Features with highest absolute correlation (originally all negative) are the average number of used cell towers in one week, distance traveled in a day of week and the average daily radius.}
\label{fig:ind_correlation}
\end{figure}

We utilize point biserial correlation to define the basic correlation between each feature and user default status due to its suitability for handling both categorical and continuous variables as are present in our feature set. This process indicates that individual features tend to have very little correlation to default status (shown in Fig. \ref{fig:ind_correlation}), with a maximum absolute correlation of only $0.1155$ and average correlation of $-0.006$. This poor individual correlation shows that no single feature reproduces the default status of the individuals reliably, proving the value of building an accurate predictive model through more complex analysis.

\subsection{Modeling}
\subsubsection{Logistic regression}
To create the predictive models, we start by defining the notation for logistic regression. 
Let $\beta_0$ denote the intercept and $\beta$ stand for a vector of regression coefficients of length $p$.
Both are determined by maximum likelihood estimation method which uses a training set with $n$ known outcomes in the
vector $Y$ for nodes in the training set. Matrix $X$ of size $n\times p$ with rows corresponding to nodes and columns corresponding to features contains values of features for all nodes in training set. For any node $u$ in test set and its corresponding vector of feature values $x_u$ of size $p$, the probability of ``success'' (in our case the default status of this node being 1) is defined as
\begin{equation}
P(y_u=1|x_u) = \frac{1}{1+\exp^{-\beta_0 - \beta\cdot x_u}},
\end{equation}
where $y_u$ denotes the value of the dependent variable for node $u$.

\subsubsection{Principal Component Analysis}
To build a more complex model that utilizes the other $6048$ features, we first normalize the features to mean values of $0$ and variance $1$. Once normalized, to deal with the extremely large number of features we perform Principal Component Analysis~\cite{pca}(PCA) to decompose the feature space and yield a set of ``principal components''. The results are linearly uncorrelated and ranked in such a way that the first principal component explains the highest possible amount of variability in the data, while each following component explains the highest amount of variance under the condition that it is orthogonal to all preceding components. Once these components are obtained, we select subsets of the components that account for large amounts of variability in the data and use them as explanatory variables for a linear model. We do this for both large and small subsets of the components, as pruning the data in this way represents a careful balance between model simplicity and accuracy. In our dataset, this balance manifests in the rapidly diminishing returns seen from larger component inclusion. The first component explains $20\%$ of variability in the data while the second explains only $7\%$. In fact, the first thirty components together explain only $42\%$ of the variability, while the first five hundred sum up to just $66\%$. For this reason, we generate two distinct logistic models: a simple one based on the first $30$ PCA components (\textit{pca-30}) and a more complex one, based on the first $500$ PCA components (\textit{pca-500}). We use notation $p1$ for the number of selected principal components. We then map our features into the new space to fit these groupings, such that
$$ X_{PC} = X \cdot C,$$
where $C\in \mathbb{R}^{p \times p_1}$ is the matrix of $p_1$ principal components and $X_{PC} \in \mathbb{R}^{n\times p_1}$ is the new feature matrix. Finally, we also fit a model (\textit{pval-05}) using only those variables from $\textit{pca-500}$ that have a $p-value<0.5$ to include the greater detail of the larger model with a lower complexity for calculation.

\subsubsection{Other models}
We further examine the data by utilizing other techniques, and build models for each as a comparison to the above described PCA models. For instance, one large issue with our dataset is that positive samples (defaulted users) account for only $0.25\%$ of the whole dataset, making it extremely unbalanced. Learning from this kind of unbalanced dataset is a well-known challenge in the data mining community that we attempt to account for by performing oversampling (multiplying positive samples to make them a larger portion of the dataset)~\cite{imbalanced}. In this case, we use multiplication factors ranging from $2-100$, but in the model benchmarks presented here, we show only the model with a multiplication factor of $2$ (\textit{oversampled-2}) since it performed the best. Oversampling was applied to the reduced dataset ($X_{PC}$). 

Further, as an alternative to the PCA reduction presented in the prior section, we also build a separate model by selecting features via Lasso regression~\cite{lasso}. This method adds a penalty term to the log likelihood function in the prediction that shrinks the coefficients of less important variables to zero. We fine tune this reduction via a free parameter $\lambda$, varying it through a range of values to obtain the best lasso fits. The efficacy of this method is demonstrated in two separate models: one based on logistic regression (\textit{lasso-logistic}) and another using a Support Vector Machine~\cite{svm} (\textit{lasso-svm}). 

Finally, the last model we use for comparison utilizes a simple method for feature space reduction that aggregates some of the more specific features into general descriptions of behavior. For instance, features that were based on a particular day or week of the three month period are grouped and combined, creating instead features for weekdays versus weekend averages or business hours versus off-hours. This aggregation not only simplifies the model, but also makes it generalizable to other datasets with different levels of detail. This method shrinks the overall feature set considerably, leaving only $781$ features. As before, these features are then normalized before being further reduced via PCA.

\section{Results and discussion}
\subsection{Experimental setting}\label{cap:exp_setting}

After each of the above models is fit to a training set comprising $70\%$ of the whole dataset, they are tested on the remaining $30\%$ of the data and evaluated via their recall, fall-out, and precision. Defaulting users are labeled as positive examples and default predictions are defined as those users within the $95^{th}$ percentile of default probabilities. This threshold is chosen to be relatively low to fit the nature of our study where false positives (non-defaulted users that were predicted as defaulters) are less damaging than false negatives (unidentified defaulters) to companies. 

The recall (the true positive rate) is the rate of identified defaulting customers. The fall-out (the false positive rate) is the probability of labeling a good client as a defaulting one. The precision is the fraction of correct default predictions out of all default predictions. These metrics are calculated as  
$$recall = \frac{TP}{TP + FN}, \quad fall\textit{-}out = \frac{FP}{FP+TN}, \quad precision = \frac{TP}{TP + FP}$$
where TP is the number of true positives, FN is the number of false negatives, FP is the number of false positive, and TN is the number of true negatives.

\subsection{Evaluation}
\begin{figure}[!ht]
  \centering
  \includegraphics[width=1\textwidth]{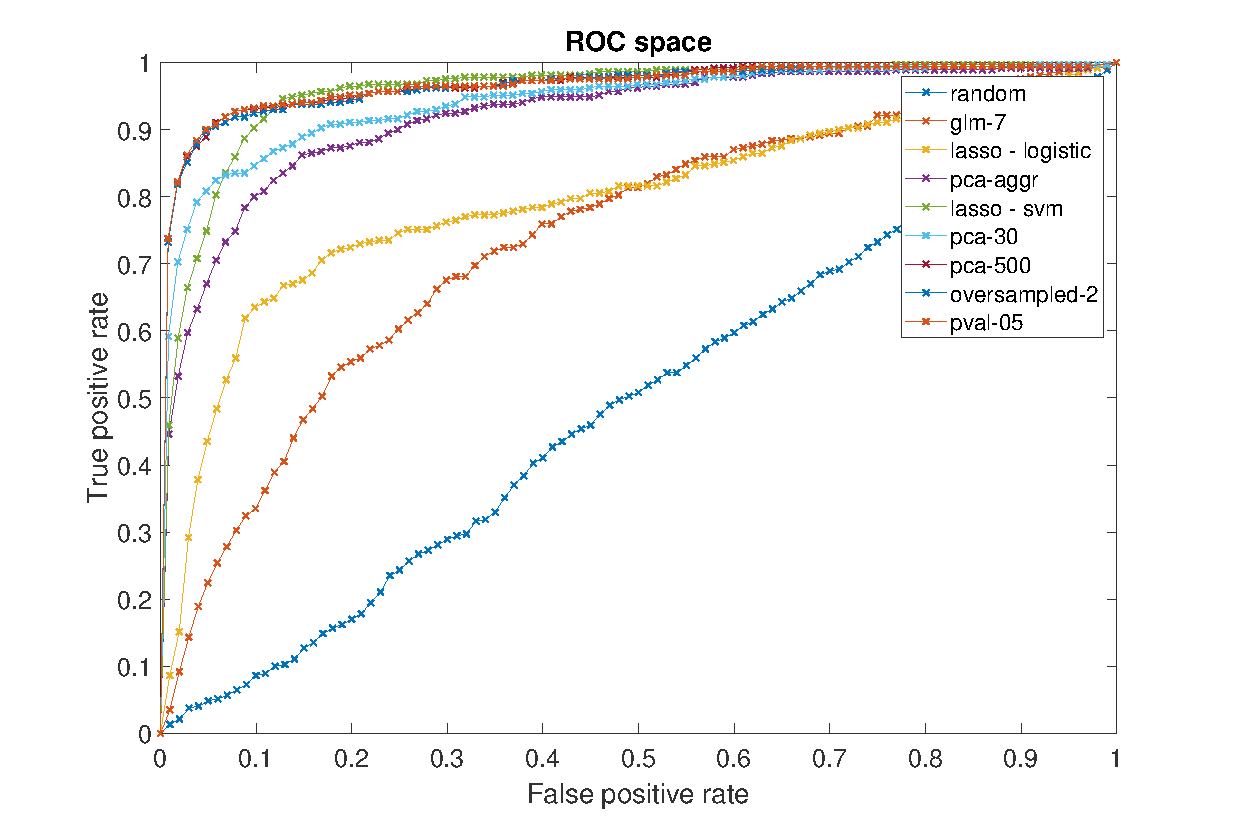}
  \caption{Model performance when predicting defaulting users from worst to best: 
  		\textit{glm-7}, \textit{lasso-logistic}, 	 
  		\textit{pca-aggr}, \textit{lasso-svm}, \textit{pca-30}, \textit{oversampled-2}, 			\textit{pca-500} and \textit{pval-500}.}
  \label{fig:model_results}
\end{figure}

The performance of the models tested can be compared using a receiver operating characteristic curve (ROC), which we show in Figure \ref{fig:model_results}. Here, we focus this comparison on the recall and fall-out of the model, where recall (which identifies defaulting customers) is of interest to phone companies, while fallout has greater relevancy in other applications of the models. The exact values of these performance metrics can be seen in Table \ref{tab:recall}, which also shows the precision of each model for reference. 

The initial results are in line with what would be expected intuitively: the worst performance comes from the random model, followed by \textit{glm-7} (the logistic model based on only $7$ features). As the models become more complicated, their performance tends to increase. For instance, a significant reduction in false positives can be seen in the \emph{lasso-logistic model} (which uses 309 variables) over the \emph{glm-7}, then a further reduction in the \emph{lasso-svm} model that uses $475$ variables. Additionally in both Lasso models, the variable with the highest coefficient is the user's most commonly utilized cell tower, indicating the presence of geographic regions which are high risk areas for user defaults. Other high performance models include the various PCA models, led by \textit{pca-500}, \textit{pval-05} and \textit{oversampled-2} with \textit{pval-05} outperforming all other models.

From these results, it is clear that for the most part larger feature sets allow for more accurate models (as would be expected), but this is not a strict rule. The best performing model is the one that begins with a large sample of features, but is then stripped of those that aren't considered significant. In other words, there are likely some 'false flags' in the feature list that tend to confuse the model rather than contribute. Feature removal only works to a point, however, as the more aggressive methods in the \emph{pca-aggr} model lose these benefits and in fact make it one of the worst performing models tested. Thus, there is likely some very meaningful information even within the extremely specific features that would be intuitively too limited to contribute much. Finally, it should be noted that while the highest performing oversampling model, \emph{oversampled-2} yields improvements comparable to filtering insignificant features, applying both techniques \emph{worsens} the results as the model apparently over-fits.

\begin{table*}[!ht]
  \centering
  \begin{tabular}{ |l|c|c|c| } 
     \hline 
     Model & Recall & Fall-out & Precision \\ 
     \hline
     random & 0.060 & 0.0501 & 0.003 \\
     glm-7 & 0.224  & 0.0495 & 0.012 \\
     lasso-logistic & 0.484 & 0.0490 & 0.023 \\
     pca-aggr & 0.676 & 0.0484 & 0.036 \\
     lasso-svm & 0.749 & 0.0482 & 0.040 \\
     pca-30 & 0.810 & 0.0480 & 0.043  \\
     pca-500 & 0.889  & 0.0478 & 0.047 \\
     oversampled-2 & 0.897 & 0.0478 & 0.047 \\
     pval-05 & {\bf 0.900} & {\bf 0.0477} & {\bf 0.048}\\
     \hline
  \end{tabular}
  \caption{Recall, fall-out and precision for each of the models presented in Figure \ref{fig:model_results}. Precision is low due to the fact that the dataset is majorly unbalanced; however, precision of the best model is about $15$ times higher than in the random model.}
  \label{tab:recall}
  \vspace{-0.3cm}
\end{table*}

\subsection{Stability}
For a more in depth look into the differences in prediction among the top three models (\emph{oversampled-2}, \emph{pca-500}, \emph{pval-05}), we look at the overlap size (the intersection of correctly labeled defaulting customers) and average overlap score (the similarity of two rankings at increasing depth giving higher weights to higher ranked observations; AOS). All analysis is restricted to only those customers with a calculated probability of default (PD) in the $95-th$ quantile. As can be seen in Table \ref{tab:stability},  both the overlap and AOS are extremely high for all model comparisons, indicating that all three models are stable and predict roughly the same nodes for default. Further, the \emph{pval-05} and \emph{pca-500} models can be seen to be more similar to each other than the \emph{oversampled-2} model due to \emph{pval-05} being essentially \emph{pca-500} with some of the problematic features removed.

\begin{table*}[!ht]
  \centering
  \begin{tabular}{ |l l|c c c| } 
     \hline 
     	& & pval-05 & pca-500 & oversampled-2 \\
     \hline
     \multirow{2}{*}{pval-05} & overlap & 100 & 98.8 & 97.3 \\
     		& AOS & 100 & 96.5 & 94.3 \\
     \hline
     \multirow{2}{*}{pca-500} & overlap & & 100 & 97.9 \\
     		& AOS & & 100 & 95.5\\
     \hline
     \multirow{2}{*}{oversampled-2} & overlap & & & 100\\
     			 & AOS & & & 100\\
     \hline
  \end{tabular}
  \caption{Overlap size (in percentage) and average overlap score of correctly labeled defaulting customers for the three best models.}
  \label{tab:stability}
  \vspace{-0.3cm}
\end{table*}

\subsection{Contribution of feature sets}
Finally, we identify the original features that contribute the most to our models to gain a better understanding of what aspects of the mobile phone data are most important for predicting default status. However, as we mentioned before in Sec. \ref{sec_correlation}, each feature is generally fairly poorly correlated with defaults. Thus it is no surprise that there is no one outstanding feature that stands out as the strongest contributor. To remedy this, we perform a general analysis using the feature set groupings described in Sec. \ref{cap:features} to identify which class of features contributes the most. By building a host of modified models, each consisting of all of the feature sets except one, we can compare the performance of the models with the reduced data against the model with all data included. This method allows us to gain a better understanding of how much information is lost when a feature set is removed (i.e. how unique that information is), data that is especially valuable to this study considering the high redundancy in our data set.

\iffalse 
\begin{table*}[t]
  \centering
  \begin{tabular}{ |l|l r|l r|l r| } 
  	\hline
    	& 
        \multicolumn{2}{ c| }{Recall} & 
        \multicolumn{2}{ c| }{Precision} & 
        \multicolumn{2}{ c| }{Contribution} \\
     Model	& glm & glm-fit & glm & glm-fit & recall & precision \\ 
     \hline
     Full feature set & 0.9  &  0.91  &  0.049  &  0.049  &   -  & -  \\ 
     - consumption f. & 0.8797 & 0.8610	& 0.0468 & 0.0458 & 0.03 & 0.002 \\ 
     - correspondent f. & 0.5757 & 0.5649 & 0.0304 & 0.0298 & {\bf 0.33}& {\bf 0.019}\\ 
     - reciprocated f. & 0.8784 & 0.8865  & 0.0464 & 0.0468 & 0.03 & 0.002 \\ 
     - mobility f. & 0.8784 & 0.8811 & 0.0464 & 0.0465 & 0.03 & 0.002 \\  
     - network f. & 0.8922 & 0.8972 & 0.0502 & 0.0481 & 0.02 & 0.001 \\  
     - cell tower PD f. & 0.9135 & 0.9054 & 0.0482 & 0.0478 & 0 & 0.001 \\ 
     \hline
     correspondent f. & 0.8640 & 0.8429 & 0.0487 & 0.0475 & - & - \\
     \hline
  \end{tabular}
  \caption{Results of models based on reduced feature sets. In each row, the underlying dataset is missing one category of features (listed in first column of the table). Two models that proved to be the best were fitted in each case: \textit{glm-500} and \textit{glm-500-filt}. Most features contribute only about $2$-$3\%$ to the final recall and $0.1$-$0.2\%$ to the final precision, except for correspondent features which contribute $33\%$ of recall and $1.9\%$ of precision.}
  \label{tab:contribution}
  \vspace{-0.3cm}
\end{table*}
\fi

\begin{table*}[t]
  \centering
  \begin{tabular}{ | l | c | c | l r | } 
  	\hline
    	 Model & Recall & Precision & $\Delta_{Recall}$ & $\Delta_{Precision}$ \\
     \hline
     Full feature set & 0.9   &  0.048  &   -  & -  \\ 
     - consumption features & 0.88 	& 0.047 & 0.02 & 0.001 \\ 
     - correspondent features & 0.58 & 0.031 & {\bf 0.32}& {\bf 0.017}\\ 
     - reciprocated features & 0.88 & 0.046 & 0.02 & 0.002 \\ 
     - mobility features & 0.88 & 0.046 & 0.02 & 0.002 \\  
     - network features & 0.88  & 0.05 & 0.02 & -0.002 \\  
     - cell tower PD features & 0.89 & 0.050 & 0.01 & -0.002 \\ 
     \hline
     Only correspondent features & 0.86 & 0.049 & - & - \\
     \hline
  \end{tabular}
  \caption{Results of models based on reduced feature sets. In each row, the underlying dataset is missing one category of features (listed in first column of the table). Two models that proved to be the best were fitted in each case: \textit{glm-500} and \textit{glm-500-filt}. Most features contribute only about $2$-$3\%$ to the final recall and $0.1$-$0.2\%$ to the final precision, except for correspondent features which contribute $33\%$ of recall and $1.9\%$ of precision.}
  \label{tab:contribution}
  \vspace{-0.3cm}
\end{table*}

As shown in Table \ref{tab:contribution}, by far the highest contributors to recall and precision are the correspondent features, which focus on individual's unique frequent correspondents in various time frames. To better understand exactly what features within the correspondent feature set are important, we perform 5-fold cross validation on a linear model using only these features. On average, this model is able to achieve a $0.86$ recall and $0.049$ precision, both fairly high for such a small subset of all features. Further, we explore which out of the $2543$ correspondent features contribute most to the final PD by mapping the maximum likelihood estimation (MLE) coefficients of PCA components back to original features' coefficients,
$$ \hat{\beta} = C \cdot \hat{\beta_{PC}} $$
where $\hat{\beta} \in \mathbb{R}^{2543\times 1} $, $C\in \mathbb{R}^{2543\times 500}$ is the matrix of principal components and $\hat{\beta}_{PC} \in \mathbb{R}^{500\times 1}$ are the MLE coefficients for a model fitted on $500$ PCA components.

We then multiply each regression coefficient by the mean of the corresponding feature (separately for paying and defaulting customers). Finally, we define a score for each user based on the values of $\hat{\beta} X_j,$ $j=1,...n$. This creates a system where higher scores correspond to customers with higher PD's, and leads to an average score of $-0.0453$ for paying customers and $16.2758$ for defaulting customers. This analysis is not possible using the whole feature set together, as the normalization of the features leads to an average contribution of zero, but the absolute contribution values of each feature are taken into account to maintain a focus on the actual impact of the features on the PD. The results for the four strongest features are presented in Table \ref{tab:corr_contribution}. In the end, the same ten features are identified as the strongest contributors for both the paying and defaulting customer groups. All ten relate to the number of unique correspondents with whom the user communicates during the holiday period around Christmas and New Year's Eve. This strongly indicates a tie between not only unique correspondents and defaulting, but also unique correspondents around holiday periods (when users are most likely to be contacting close family and friends). From these results we can begin to draw a clear connection of low probability of default with the large number of active unique ties an individual has within the network, and the high strength of those ties.

\begin{table*}[t]
  \centering
\begin{tabular}{ | l | c | } 
  	\hline
     Feature	& Mean relative contribution  \\ 
     \hline
     unique message correspondents (12/24) &  0.043 \\ 
     unique message correspondents (incoming, 12/24)  & 0.041 \\ 
     unique message correspondents (outgoing, 12/24)  & 0.024 \\ 
     unique correspondents (outgoing, 12/24)  & 0.023\\ 
     \hline
     sum & 0.131  \\
     \hline
  \end{tabular}
  \caption{Four features with the highest contribution to the PD calculated on separate sets of paying and defaulting customers. All of the top $10$ relate to the number of unique correspondents during holiday period around Christmas and New Year's Eve.}
  \label{tab:corr_contribution}
  \vspace{-0.3cm}
\end{table*}

\iffalse
\begin{table*}[t]
  \centering
\begin{tabular}{ |l|l r|l r| } 
  	\hline
    	& 
        \multicolumn{2}{ c| }{Mean Contr.} & 
        \multicolumn{2}{ c| }{Mean rel. contr.}  \\
     Feature	& pay. cst & def. cst & pay. cst & def. cst  \\ 
     \hline
     uniqCorrDay20161224DataInOut & -0.0020 & 0.7058 & 0.0434 & 0.0434 \\ 
     uniqCorrDay20161224DataIn & -0.0018 & 0.6588 & 0.0405 & 0.0405 \\ 
     uniqCorrDay20161224DataOut & -0.0011  & 0.3913 & 0.0240 & 0.0240 \\ 
     uniqCorrDay20161224AllOut & -0.0010 & 0.3688 & 0.0227 & 0.0227\\ 
     uniqCorrDay20161224AllInOut & -0.0010 & 0.3676 & 0.0226 & 0.0226 \\  
     uniqCorrDay20161227VoiceOut & -0.0010 & 0.3560	& 0.0219 & 0.0219 \\  
     uniqCorrWeek20161226VoiceOut & -0.0010 & 0.3444 & 0.0212 & 0.0212 \\ 
     uniqCorrWeek20161226AllOut & -0.0009 &  0.3406	& 0.0209 & 0.0209 \\
	 uniqCorrDay20161224AllIn & -0.0009 & 0.3272	& 0.0201 & 0.0201 \\
	 uniqCorrDay20161231AllOut & -0.0009 & 0.3196 & 0.0196 & 0.0196 \\
     \hline
     sum & -0.0116 & 4.1800	& 0.2568 & 0.2568 \\
     \hline
  \end{tabular}
  \caption{Ten features with the highest contribution to the PD calculated on separate sets of paying and defaulting customers. All of them relate to the number of unique correspondents during holiday period around Christmas and New Year's Eve.}
  \label{tab:corr_contribution}
  \vspace{-0.3cm}
\end{table*}
\fi

However, it should be noted that this kind of estimation of variable contribution is too simple to provide a definitive insight into the relative importance of the variables. There are several other approaches that compare predictors in regression including methods considering variable importance via $R^2$ partitions~\cite{pratt1987,thomas2008measuring}, dominance analysis~\cite{azen2003dominance} and relative weights analysis~\cite{johnson2000heuristic,lebreton2008multivariate}. According to the literature, the dominance analysis provides the most accurate results, but it is not practical for our problem since it measures relative importance in a pairwise fashion and is suitable only for small variable sets. Instead, the relative weights approach is sometimes used for systems with multiple correlated predictors as can be found here~\cite{lebreton2008multivariate}, but this approach is theoretically flawed and is therefore not recommended for use~\cite{thomas2014johnson}. We instead utilize a variable importance (VI) extension for logistic regression~\citep{thomas2008measuring} that is based on Pratt's axiomatic~\citep{pratt1987} and the geometric approach of Thomas~\citep{thomas1998variable}. This method equates VI to variance explained by the variable, which is $\beta_j\rho_j$, where $\beta_j$ is the standardized regression coefficient and $\rho_j$ is the simple correlation between variables $Y$ and $X_j$. For our purposes, we use the geometric approach to this method; an interpretation of Pratt's measure based on the geometry of least squares. Here, the VI indices are defined as
\begin{equation}
d_j = \frac{\hat{\beta}_j\hat{\rho_j}}{R^2}, j=1,\dots, p,
\end{equation}
where hats denote sample estimates, and $R^2$ is as usual the proportion of sample variance explained.

Further, we utilize a pseudo-$R^2$ measure based on Weighted Least Squares such that this set of indices sums to one and the importance of a subset of variables is equal to the sum of their individual importance. This results in the four most important features, according to VI metric, shown in Table \ref{tab:d_WLS}, yielding results very similar to the ones presented in Table \ref{tab:corr_contribution}. The two most important variables are: the number of unique correspondents with whom the user had a call on Dec 24th with relative importance of $0.048$, and the number of unique correspondents from which the user received a message on Dec 24th with relative importance of $0.045$. This finding reiterates the significance of the number of unique correspondents
with whom the user communicates around holiday periods. Finally, to ensure that these features actually carry unique information, we fit another model removing from the dataset the features with the highest identified VI. Doing so reduces the recall by $0.042$ and leaves none of the new identified important variables with a relative importance above $0.03$.

\begin{table*}[t]
  \centering
\begin{tabular}{ | l | l | } 
     \hline
     Feature	& $d_j$ \\ 
     \hline
     unique message correspondents (12/24) & 0.048 \\ 
     unique message correspondents (incoming, 12/24) & 0.045 \\ 
     unique message correspondents (outgoing, 12/24) & 0.029 \\ 
     unique correspondents (outgoing, 12/24) & 0.026\\ 
     \hline
     sum & 0.148 \\
     \hline
  \end{tabular}
  \caption{Ten features with the highest contribution to the PD. All of them relate to the number of unique correspondents during holiday period around Christmas and New Year's Eve.}
  \label{tab:d_WLS}
  \vspace{-0.3cm}
\end{table*}

\iffalse
\begin{table*}[t]
  \centering
\begin{tabular}{ | l | l | } 
     \hline
     Feature	& $d_j$ \\ 
     \hline
     uniqCorrDay20161224DataInOut & 0.0482 \\ 
     uniqCorrDay20161224DataIn & 0.0447 \\ 
     uniqCorrDay20161224DataOut & 0.0288 \\ 
     uniqCorrDay20161224AllInOut & 0.0260\\ 
     uniqCorrDay20161224AllOut & 0.0257 \\  
     uniqCorrDay20161227VoiceOut & 0.0243 \\  
     uniqCorrDay20161224AllIn & 0.0235 \\ 
     uniqCorrWeek20161226VoiceOut & 0.0231 \\
	 uniqCorrDay20161226AllOut & 0.0230 \\
	 uniqCorrDay20161231AllOut & 0.0218 \\
     \hline
     sum & 0.2890 \\
     \hline
  \end{tabular}
  \caption{Ten features with the highest contribution to the PD. All of them relate to the number of unique correspondents during holiday period around Christmas and New Year's Eve.}
  \label{tab:d_WLS}
  \vspace{-0.3cm}
\end{table*}
\fi

\iffalse
\begin{table*}[t]
  \centering
\begin{tabular}{ | l | l | } 
     \hline
     Feature	& $d_j$ \\ 
     \hline
     nRecsDataInOut	& 0.0282 \\
	uniqCorrHourOfWeek615VoiceInOut	& 0.0274 \\
	nContactsData & 0.0264	\\
	nContactsFreqVoice & 0.0252	\\
 	nContactsAll	& 0.0226	\\
 	uniqCorrTotalAllInOut & 0.0214	\\
	nContactsFreqData & 0.0213	\\
	uniqCorrRestDataIn & 0.0207	\\
	uniqCorrOfficeAllOut & 0.0190	\\
	uniqCorrTotalVoiceOut & 0.0187	\\
     \hline
     sum & 0.2309 \\
     \hline
  \end{tabular}
  \caption{}
  \label{tab:d_WLS_red}
  \vspace{-0.3cm}
\end{table*}
\fi

\section{Conclusion}
In this paper, we investigate many different aspects of user behavior to build a large suite of features for analysis, starting with constructing the underlying social networks based on the cell phone usage data. This produces a complex network rich with information. Applying common network metrics reveals some of the characteristics of the network such as the heterogeneity of the centrality, diameter, and reciprocity measures.

The complex nature of the problem addressed by our machine learning method to predict whether users will default in paying their cellphone bill  leads us to utilize some $6000$ features, which we then pare down to only the most predictive ones. 
The resulting model achieves a recall of $0.9$ with a fall-out of only $0.048$, the performance that compares favorably
with \cite{san2015mobiscore,agarwal2018predicting}, e.g., recall of $0.674$ is reported in this reference.  

Finally by investigating in depth the various features that contribute to the model, we are able to pinpoint the surprisingly best, contributor: the number of unique contacts with whom the user interacted around the winter holidays (when users are most likely to contact their closest friends and family). 
The significance of this correspondent information is higher than that of more traditionally used features. These results demonstrate the need for systematic approach to selecting features for complex behavior prediction. Indeed, our results show that the strength of links within the network is better determined by the timing of communications rather than the volume, duration, or distance traditionally used for the similar predictions.

While our use of CDR data here focuses on user analytics and predictive modeling, CDR datasets can support many other avenues of research. This work investigates the default status of individuals, but there are many other complex aspects of user behavior that could benefit from similar computational techniques. The modeling we present could be improved by gathering information over larger time periods in order to get a larger population of defaulting individuals. 
Finally, with new machine learning algorithms constantly being designed and improved, a more specific algorithm that is built for high dimensional data such as ours could improve the resultant predictions and understanding of how the specific features contribute to the overall model which is novel and unexpected. 
Hence, our novel approach leading to these results is the main contribution of our paper to the state of the art.

\section{Acknowledgments}
Funding: This work was supported in part by the Army Research Laboratory (ARL) under Cooperative Agreement Number W911NF-09-2-0053 (NS-CTA), by the Office of Naval Research (ONR) Grant No. N00014-15-1-2640, by the RENOIR EU H2020 project under the Marie Sk\l{}odowska-Curie Grant Agreement No. 691152 and by the National Science Centre, Poland, project no. 2016/21/B/ST6/01463. The views and conclusions contained in this document are those of the authors and should not be interpreted as representing the official policies either expressed or implied of the Army Research Laboratory or the U.S. Government. 

\section*{References}

\bibliography{bibliography}

\appendix

\section{Constructing an unweighted network representation with frequency cutoffs}
\label{unweighted_network}

We define the unweighted network via a cutoff, $c$, representing the minimum number of communications from node $i$ to node $j$ required for an edge to be drawn ~\cite{Newman2001}. Specifically, if you consider the weight of communications between node $i$ and node $j$ to be the total volume of communications $w_{i,j}$, then the condition for a directional link being formed from node $i$ to node $j$ is $w_{i,j} \geq c$. This link then becomes an unweighted directional edge from node $i$ to node $j$ without restricting the possibility of another link forming from node $j$ to node $i$ to create an unweighted bidirectional edge. One main advantage of constructing the network in this way is that it allows the network to be simplified, removing noise while still retaining information regarding the frequency of communication. Still, the methodology proposed here contains some inherent loss of information partially stemming from the use of the raw number of communications between individuals. To account for possible differences in the amount of information contained in calls versus texts, it is natural to attempt to use different communication weights based on communication type. As such, we experimented with such methods to better account for these information exchange levels and found that any such scheme would cause bias between age groups. In general, since younger individuals used more text messages while older individuals used more calls we treat their contributions to the network flatly to minimize bias.

\begin{figure}[t]
\centering
\includegraphics[width=1\textwidth]{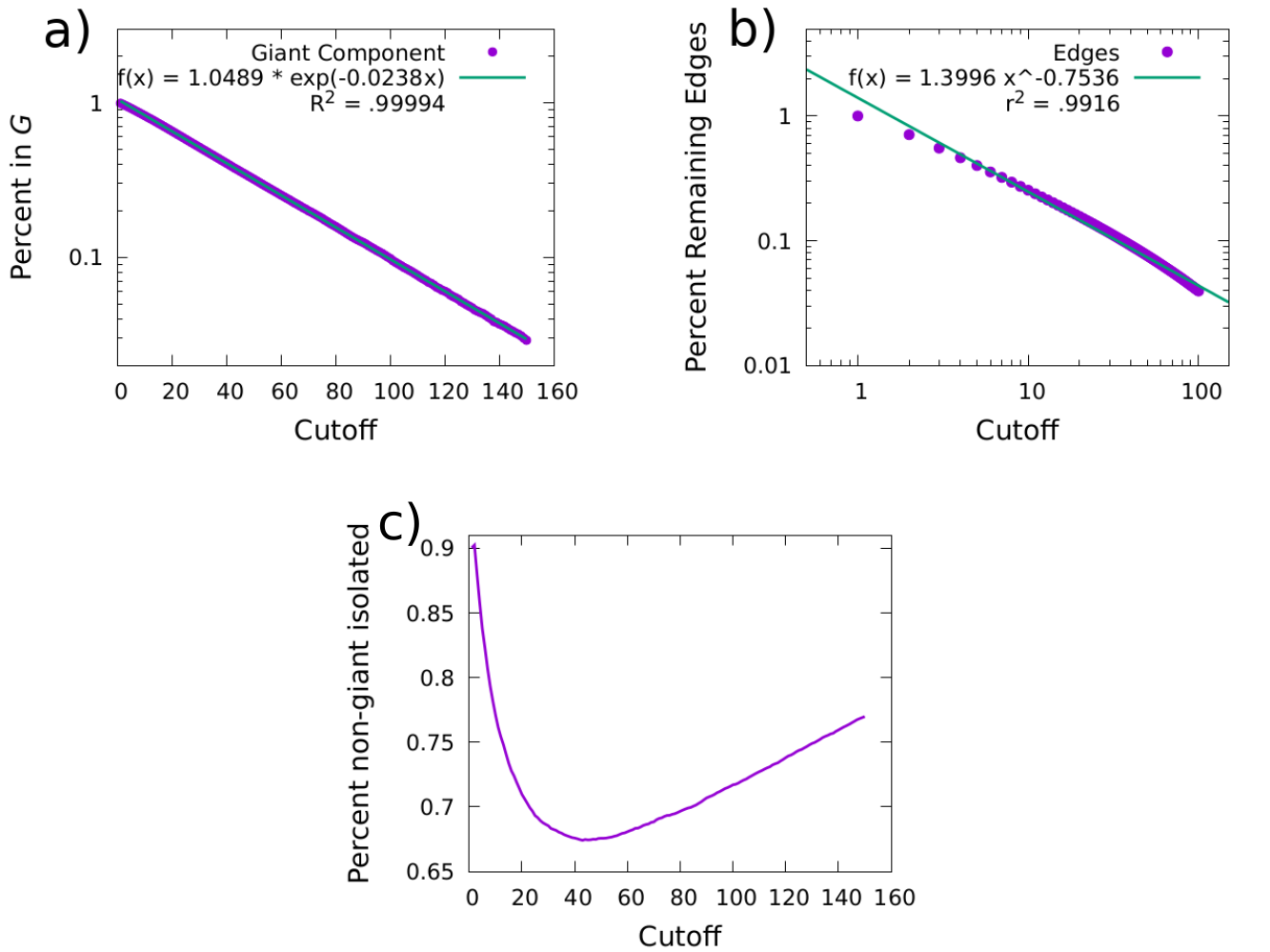}
\caption{(a) The giant component of the social network decays exponentially ($\lambda=0.0238$) with increased minimum number of communications required for an edge to be drawn. (b) The number edges in the network also decays rapidly with increased cutoff, closely fitting a power law with $\gamma=0.7536$. (c) The percentage of non-giant component nodes that are isolated for a given cutoff. The isolated nodes reach a minimum at a cutoff of $43$.}
\label{cutoff_decay}
\end{figure}

This method yields a strict structural view of the true friendship network, addressing the need for noise reduction by pruning down the network as well as showing how robust the network is to increasingly strict friendship requirements.~\citep{Yan2018} In Fig. \ref{cutoff_decay}(a), we show that the giant component of the network decays exponentially with an increased frequency cutoff for edges, and is generally very sensitive to such an increase. Without a cutoff, the giant component of the network contains $99.1\%$ of the nodes, but it drops to only half of the network when the frequency cutoff reaches $30$. A similar effect can be seen in Fig.\ref{cutoff_decay}(b), where the total number of edges in the social network shows a power law decay for increasing cutoff values. This power law decay of edges implies that the scheme is useful for removing noisy, low frequency communication while leaving intact the dense communications representative of strong social ties. The proper cutoff value for a given situation is difficult to establish, however, since the vast majority of connections in the network are low frequency. In fact, less than half of the original edges remain when a cutoff of $c=4$ is implemented. As the cutoff increases further, the network loses much of its connectivity and instead shows a rough sketch of the community structure of the system. This effect is seen in Fig. \ref{cutoff_decay}(c), in which the percentage of nodes outside of the giant component that are isolated \emph{decreases} with increased cutoff, the result of small communities being separated from the giant component but remaining intra-connected. Of course, these results yield only a crude approximation of the community structure; high cutoffs lower the connectivity within communities as well as separate them from the giant component. This trade-off becomes apparent at cutoffs larger than $c=43$, where the number of isolated nodes increases and even tightly bound communities unravel. A more direct and robust detection of the network's community structure is discussed in Sec. \ref{communities}.

\begin{figure}[t]
\centering
\includegraphics[width=.75\textwidth]{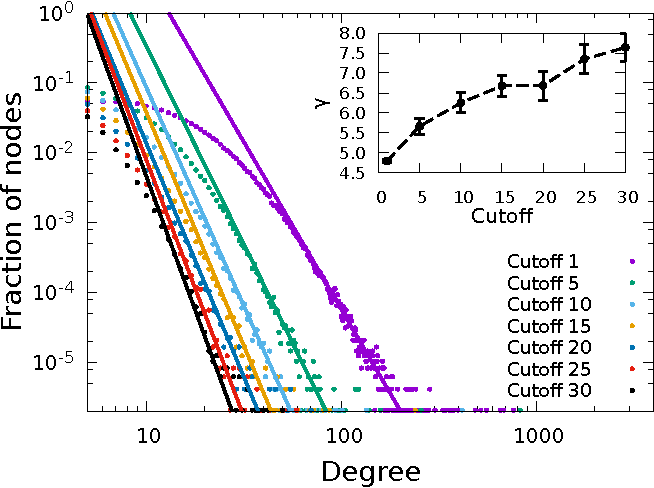}
\caption{The out-degree distribution of the mobile phone network for various edge weight cutoffs. The inset shows the estimated power law exponent for the various cutoff values to highlight the different scaling rates.}
\label{degree_dist}
\end{figure}

In addition to the effects on the connectivity of the network, the frequency cutoff also has a significant impact on the overall degree distribution (shown in Fig. \ref{degree_dist}). As expected~\cite{Li2014,Onnela2007,Barabasi2016}, for all values of the cutoff the degree distribution shows a power law tail. However, the rate at which the tail decays changes as the edges are removed. As the cutoff increases, many of the highest degree nodes lose the vast majority of their edges. As a result, the higher cutoffs greatly reduce the maximum degree of the network and increase the power scaling exponent of the degree distribution causing faster tail decay~\cite{Alstott2014, Klaus2011, Clauset2009}. As with the prior results on increased cutoff, however, this is only true to a point. The exponent increase appears to saturate at high values of the cutoff where the impact gets weaker and weaker. These properties allow us to use an unweighted graph with low cutoffs to define basic network features such as node degree, while higher cutoff values are useful for finding the stable number of strong contacts.

\section{Reciprocity Measures}
\label{app:reciprocity}
Here we generalize how reciprocity is defined using $R_i=\sum_j {N_{i,j}r_{i,j}}$ where $r_{i,j}$ is the reciprocity score between node $i$ and $j$ and $N_{i,j}$ is the weight of that score. To keep the metric bounds at $[-1,1]$, we constrain $r_{i,j}$ to $[-1,1]$ and require $N_{i,j}$ to be normalized, ($1=\sum_j N_{i,j}$). Previously, we defined a metric using $r_{i,j}=\frac{w_{i,j}-w_{j,i}}{w_{i,j}+w_{j,i}}$ and where all scores are equal. 
Under normalization this means that $N_{i,j}= \frac{1}{k_i}$ where $k_i$ is the total degree of node $i$. Here we introduce two variants to produce two alternative metrics for reciprocity.

First we remove the dependence on the number of communications by modifying $r_{i,j}$ to take on only three values $-1,0,1$, causing links with small $w_{i,j}$ to become more important. If between node $i$ and node $j$ there exists any number of communications from $i$ to $j$, but no communications from $j$ to $i$, then $r_{i,j}=1$. If the opposite is true, then $r_{i,j}=-1$. Finally if both $i$ and $j$ send any number of communications to each other, $r_{i,j}=0$. $N_{i,j}$ in this metric is unchanged. We call this metric binary weighted reciprocity.

The second metric modifies $N_{i,j}$ rather than $r_{i,j}$. To make high volume links more important, we set $N_{i,j}\propto w_{i,j}+w_{j,i}$. Applying normalization leads to $N_{i,j}=\frac{w_{i,j}+w_{j,i}}{\sum_{j'} w_{i,j'}+w_{j',i}}$. The simplified expression is then $R_i=\sum_{j} \frac{w_{i,j}-w_{j,i}}{\sum_{j'} w_{i,j'}+w_{j',i}}$, which we call hyper-weighted reciprocity.

\begin{figure}
\centering
\includegraphics[width=1\textwidth]{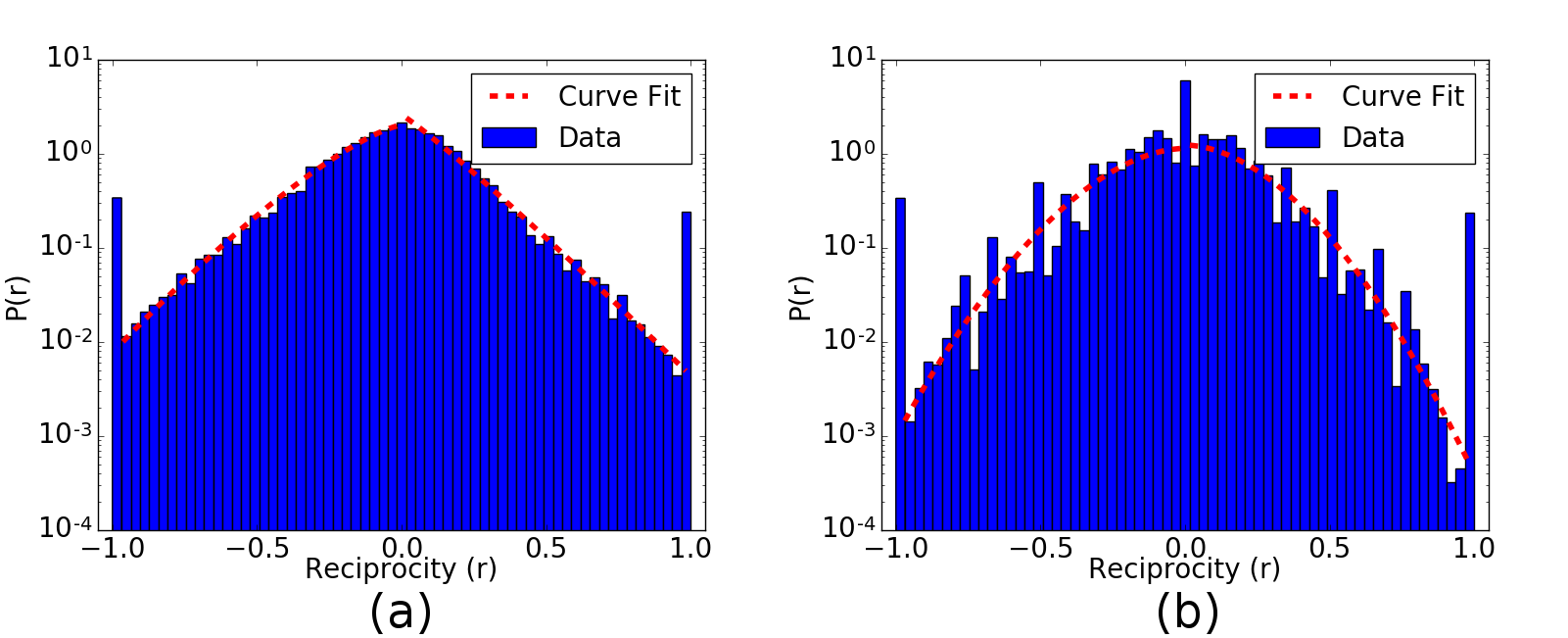}
\caption{(a) The distribution of reciprocities using the hyper-weighted metric. (b) The distribution of reciprocities using the binary metric. Both metrics maintain the extreme values. They also maintain a bell shape curve, but the binary metric exhibits multiple discrete peaks.}
\label{fig:reciprocity hyper binary}
\end{figure}

In Fig \ref{fig:reciprocity hyper binary}, we show the distribution for both of these metrics. The hyper-weighted reciprocity is very similar to the weighted reciprocity, but the binary distribution is significantly different. Most of these difference arise from the binary nature of the metric, where for example a node with degree three can only have reciprocity $0, \pm 1/3, \pm 2/3, \pm 1$. Because the system is dominated by low degree nodes, this type of scenario is frequent causing a few discrete fractions to be most prevalent. Further, we include a fit following the same methodology as in Fig \ref{fig:weighted reciprocity}, where Table \ref{tab:app_weighted_recip} shows the fitting parameters.

\begin{table*}[!ht]
  \centering
  \begin{tabular}{ |c|c|c|c|c| } 
     \hline 
     Fitting Parameter & \multicolumn{2}{c}{Binary Weighted} & \multicolumn{2}{c}{Hyper-weighted} \vline \\ 
     \hline
      & Positive & Negative & Positive & Negative \\
     \hline
     A & 0.218 & 0.136 & 0.957 & 0.707 \\
     B & 1.84  & 1.83 & 1.08 & 1,33 \\
     C & 0.124 & 0.142 & 0.157 & 0.179 \\
     $R^2$ & .798 & 0.785 & 0.957 & .976 \\
     \hline
  \end{tabular}
  \caption{The fitting parameters used in Fig \ref{fig:reciprocity hyper binary}. Positive and Negative refer to the fit of the distribution where the reciprocity values are positive and negative respectively. The Binary metric has relatively poor $R^2$ values due to the many peaks present, but the fit captures the general shape. The Hyper-weighted metric preforms better on goodness of fit and clearly shows the asymmetry between the positive and negative side of the distribution.}
  \label{tab:app_weighted_recip}
  \vspace{-0.3cm}
\end{table*}

Unfortunately, these features are not sufficiently unique from that shown in the main text of this work, and thus not suitable as separate features. The binary and weighted metrics have a Pearson correlation coefficient of $0.9224$, while the hyper-weighted and weighted metrics have a coefficient of $0.727$. Because of these high levels of correlation and despite weighting reciprocity differently, we conclude that the alternate methods do not produce any sufficiently unique information. They can, however, be substituted for the metric used based on specific use cases where the intuition behind the calculations fits better, and thus may be useful for future work.

\section{Feature Sets}
\label{app:feature_sets}
\subsection{Consumption features}
Consumption features include information about the total number of communication events, the total and average duration of phone calls, and average time between consecutive communications. These broad categories are then broken up into time windows (specific times of day or days of the week) and separated by communication type and direction to form a large contribution to our overall feature set. These features characterize the normal activity patterns for individuals within the network. On average, per day each user makes $5.3$ calls lasting about $40$ seconds each, and sends $10.4$ messages. Additionally, this feature shows a slightly higher wait time for incoming events ($3$ hours) than outgoing events ($2.5$ hours).

\subsection{Correspondent features}
Correspondent features are based on the distinct number of individuals that each user communicates with over specific time periods, separated between calls and text messages. This includes the total number of unique people that exchange at least five messages with the user on average every week, and the number of regular correspondents that make at least two calls to the user on average in a week. Further, this feature includes the percentage of incoming communications that are messages and the number of unique correspondents over various time periods and scales (such as normal business hours versus rest hours and week days versus weekends). Analysis of this feature set shows that the average number of correspondents (call or message) is $31$, while call correspondents make on average $21$ calls and message correspondents sent on average $17$ messages. However for regular correspondents, each user makes only $3.3$ calls and sends $1.7$ messages. The number of unique correspondents during the weekend drops to about half of the number for weekdays.

\subsection{Reciprocated event features} 
We define reciprocated events as events where the observed user returns a call/message within an hour of receiving a communication. We extract this information for each user and every hour of the week, as well as the median time between reciprocated call events and median time to answer messages. Similar to above, we also aggregate these indicators to days of week and time of day. On average, this analysis shows that users reciprocate only $14$ calls and $15$ messages for the night time (midnight-6am) during the whole 3-month period. Interestingly, the average time to answer a message is only $21$ minutes in the morning (6am-10am). This drops even further to a little less than $10$ minutes during other times of the day (10am-6am). Further, the average number of reciprocated messages increases as the week goes by, going from $6.5$ to $6.9$ from Monday to Friday, but then dropping to $5.7$ and $4.8$ on Saturday and Sunday, respectively.

\subsection{Mobility features} 
Mobility features describe daily movement patterns of a person. They use the cell tower location to which the individual connected to for each event. Features include average daily radius of gyration (minimum radius that encompasses all the locations visited by a user), average distance traveled per day of week (calculated as sum of distances between consecutive antennas), popular cell towers (IDs of most popular cell towers that sum up to $90\%$ of records) and average number of unique cell towers per week. A typical user's average daily radius of gyration is $8.5$ km. Interestingly, there is an order of magnitude difference when comparing mean and median distance traveled by day of the week, but the trend is the same in both cases. From Monday to Thursday, the median distance traveled is fairly constant around $33$ km, but increases to $40$ km on Friday and $47$km on Saturday before dropping to $23$ km on Sunday. Lastly, the average number of utilized cell towers per week is $15$.

\subsection{Location features}
From information about the two most commonly used cell towers and the default status of each user, we can develop location features to identify high risk geographic regions. We use two cell towers instead of one to account for users spending a large amount of time at both their residence and workplace/school. For this reason, two new features are added for each user; one for each tower. From this information, we can calculate the empirical probability of default for each cell tower, yielding a mean of $0.0032$. Note that when used for prediction purposes, these calculations are based only on the training data set and then applied to the test dataset accordingly (more information on this experimental setting is in \ref{cap:exp_setting}).

\end{document}